\documentclass{article}

\usepackage{microtype}
\usepackage{graphicx}
\usepackage{subfigure}
\usepackage{booktabs} 
\usepackage{multirow}

\usepackage{amsmath}
\usepackage{amsthm}
\usepackage{amsfonts}
\usepackage{natbib}
\usepackage{multirow}
\usepackage{soul}
\usepackage{threeparttable}

\usepackage{hyperref}

\usepackage[accepted]{icml2018}

\icmltitlerunning{Efficient Neural Audio Synthesis}

\renewcommand{\vec}{\mathbf}

\begin{document}

\twocolumn[
\icmltitle{Efficient Neural Audio Synthesis}

\icmlsetsymbol{equal}{*}

\begin{icmlauthorlist}
\icmlauthor{Nal Kalchbrenner}{equal,dm}
\icmlauthor{Erich Elsen}{equal,brain}
\icmlauthor{Karen Simonyan}{dm}
\icmlauthor{Seb Noury}{dm}
\icmlauthor{Norman Casagrande}{dm}
\icmlauthor{Edward Lockhart}{dm}  \\
\icmlauthor{Florian Stimberg}{dm}   
\icmlauthor{A\"aron van den Oord}{dm}
\icmlauthor{Sander Dieleman}{dm} 
\icmlauthor{Koray Kavukcuoglu}{dm}
\end{icmlauthorlist}

\icmlaffiliation{dm}{DeepMind}
\icmlaffiliation{brain}{Google Brain}

\icmlcorrespondingauthor{Nal Kalchbrenner}{nalk@google.com}

\icmlkeywords{Machine Learning, ICML}

\vskip 0.3in
]



\printAffiliationsAndNotice{\icmlEqualContribution} 

\begin{abstract}
Sequential models achieve state-of-the-art results in audio, visual and textual domains with respect to both estimating the data distribution and generating high-quality samples. Efficient sampling for this class of models 
has however remained an elusive problem. With a focus on text-to-speech synthesis, we describe a set of general techniques for reducing sampling time while maintaining high output quality. We first describe a single-layer recurrent neural network, the WaveRNN, with a  dual softmax layer that matches the quality of the state-of-the-art WaveNet model. The compact form of the network makes it possible to generate 24~kHz 16-bit audio 4$\times$ faster than real time on a GPU.
Second, we apply a weight pruning technique to reduce the number of weights in the WaveRNN. We find that, for a constant number of parameters, large sparse networks perform better than small dense networks and this relationship holds for sparsity levels beyond 96\%. The small number of weights in a Sparse WaveRNN makes it possible to sample high-fidelity audio on a mobile CPU in real time.
Finally, we propose a new generation scheme based on subscaling that folds a long sequence into a batch of shorter sequences and allows one to generate multiple samples at once. The Subscale WaveRNN produces 16 samples per step without loss of quality and offers an orthogonal method for increasing sampling efficiency.
\end{abstract}

\section{Introduction}
\label{submission}

Sequential generative models achieve state-of-the-art performance in a variety of domains including natural language \cite{wu,transformer}, natural images~\cite{van2016pixel, DBLP:conf/icml/ReedOKCWCBF17} and videos~\cite{DBLP:conf/icml/KalchbrennerOSD17} and speech and music~\cite{wavenet, samplernn, performance-rnn-2017, nsynth}. The models learn the joint probability of the data by factorizing the distribution into a product of conditional probabilities over each sample. This structure lets the models allot significant capacity to estimate each conditional factor, makes them robust during training and easy to evaluate. The ordering encoded in the structure also makes the sampling process strictly serial: a sample can be generated only after samples on which it depends have been produced in accordance with the ordering. The serial aspect of the sampling process can make it slow and impractical to use these models to generate high-dimensional data like speech and video. 

Our goal is to increase the efficiency of sampling from sequential models without compromising their quality. The time $T(\vec{u})$ that the sampling process takes is the product of the number of samples in the target $\vec{u}$ (e.g. the number of audio samples in a spoken utterance or the number of pixels in an image) and the time required to produce each sample. The latter can be decomposed into computation time $c(op_i)$ and overhead $d(op_i)$ for each of the $N$ layers (operations) of the model:

\begin{equation}
\label{latency}
T(\vec{u}) = |\vec{u}| \sum_{i=1}^{N} (c(op_i) + d(op_i))
\end{equation}
The value of $T(\vec{u})$ can grow prohibitively large under any of the following conditions: if $|\vec{u}|$ is large as in the case of high-fidelity audio composed of 24,000 16-bit samples per second; if $N$ is large due to the use of a very deep architecture such as WaveNet~\cite{wavenet}; if $c(op_i)$ is large due to e.g. especially wide layers or a large number of parameters; or if the overhead $d(op_i)$ is high due to the cost of launching each individual operation.

With a focus on text-to-speech synthesis, we propose a set of methods to make sampling orders of magnitude faster.
We reduce the contributions from each of the factors $N$, $d(op_i)$, $c(op_i)$, and $|\vec{u}|$ with minimal loss to the quality of the generated output.
We benchmark all models on a 
single-speaker North-American English text-to-speech
dataset where the input is composed of predicted linguistic feature vectors and the output is the raw 24 kHz, 16-bit waveform  (Section~\ref{experiments}). We report the Negative Log-Likelihood (NLL) reached by a model on held-out data, the results of A/B comparison tests between a pair of models as rated by human listeners and Mean Opinion Scores (MOS) for the samples of a model.


We begin by designing a sequence model that requires a low number $N$ of operations per sample. We make use of the core property of recurrent neural networks (RNN) that a single recurrent layer applied to the previous state can deliver a highly non-linear transformation of the context. The WaveRNN model is a single-layer RNN with a dual softmax layer that is designed to efficiently predict 16-bit raw audio samples. We see that the \mbox{WaveRNN} with 896 units achieves NLL scores comparable to those of the largest WaveNet model, there is no significant difference in audio fidelity according to a A/B comparison test (Table~\ref{ab}), and the MOS is similarly high.  
The \mbox{WaveRNN} achieves this performance by requiring just $N=5$ matrix-vector products in sequence for each 16-bit sample; for simplicity we exclude non-linearities and other minor operations from the count $N$.
This is in contrast with WaveNet that has 30 residual blocks of two layers each requiring a series of $N=30*2=60$ matrix-vector products. 

Even with the low $N$, the overhead $d(op_i)$ can still represent a significant bottleneck in a regular implementation of sampling from the WaveRNN. We sidestep the overhead by implementing custom GPU operations \cite{persistentrnns} for the sampling process. This allows the WaveRNN to generate 96,000 16-bit samples per second on a Nvidia P100 GPU, which corresponds to 4$\times$ real time of high-fidelity 24kHz 16-bit audio. As a comparison, our best GPU kernel for the WaveNet model runs at roughly 0.3$\times$ real time on the same platform. Throughput increases with a batch of 4 where the kernels achieve 39,0000 samples per second (a total throughput of 156,000 samples/sec.)

Reducing the number of parameters in the network decreases  the amount of computation $c(op_i)$ required for sampling. 
With that in mind, we aim at maximizing the performance we can get from a given amount of parameters. ~\cite{morphnet} also consider the problem of maximizing performance under a given compute budget and solve it with an approach based on neuron pruning.  We sparsify the weights in the WaveRNN using the weight pruning techniques of ~\cite{exploringsparsity,topruneornot}.
For a fixed parameter count, we discover that large sparse WaveRNNs significantly outperform small dense \mbox{WaveRNNs} and that this relationship holds up to high levels of sparsity greater than $96\%$ (Figure~\ref{fig:sparsity}).

The combination of Sparse WaveRNN's high quality output, its small number of parameters and the low requirements on memory bandwidth makes the model well-suited for efficient implementations on low-power mobile platforms (such as those found in mobile phones).
We implement and benchmark the sparse matrix-vector products and non-linearities used in the WaveRNN on a mobile CPU (Table~\ref{swrnnspeed}). 
Even though the amounts of computation and memory bandwidth are, respectively, three and two orders of magnitude smaller on a mobile CPU than on a GPU, our benchmarks on off-the-shelf mobile CPUs indicate that the resources are sufficient for real-time on-device audio synthesis with a high-quality Sparse WaveRNN. 
To our knowledge, this is the first sequential neural model  capable of real-time audio synthesis on a broad set of computing platforms including off-the-shelf mobile CPUs.

\begin{figure}
\centering
\includegraphics[scale=0.5]{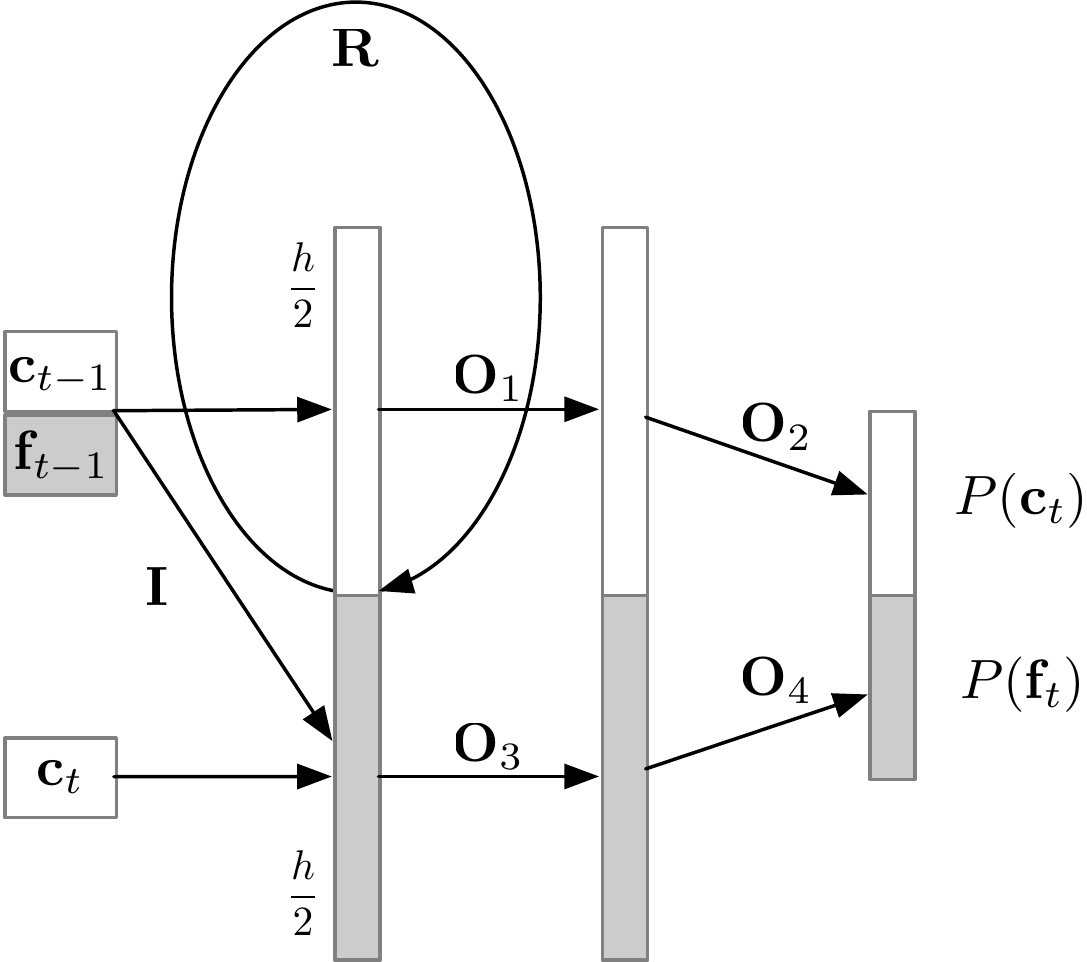}
\caption{The architecture of the WaveRNN with the dual softmax layer. $\mathbf{c}$ represents the coarse (high 8-bits) of the sample and $\mathbf{f}$ represents the fine (low 8-bits) of the sample.  The multiplication by $\mathbf{R}$ happens for both the coarse and fine bits simultaneously, then output of the gates is evaluated for the coarse bits only and $\mathbf{c}_t$ is sampled.  Once $\mathbf{c}_t$ has been sampled from $P(\vec{c}_t)$, the gates are evaluated for the fine bits and $\mathbf{f}_t$ is sampled.}
\label{wavernn}
\vskip -0.1in
\end{figure}

Finally, we tackle the contribution from the component $|\vec{u}|$ in Equation~\ref{latency}. Multiple recent approaches have the goal of making sampling from sequential models more parallel \cite{DBLP:conf/icml/ReedOKCWCBF17,DBLP:journals/corr/abs-1711-02281,parallel_wavenet}. However, these models either make local independence assumptions between generated samples undermining the backbone of sequential models, or they require training multiple domain-specific networks with specialized losses that restrict the overall usability of the models. 

\begin{table*}[t]
\vskip 0.15in
\begin{center}
\begin{small}
\begin{sc}
\begin{tabular}{lcccccccc}
\toprule

Model (vs WaveRNN-896)   & Better & Neutral & Worse & Overall & Significant  \\
\midrule
WaveNet 512 (60)   & 145 & 529 & 126 & $0.02 \pm 0.08$ & No \\
Sparse WR 384 (2048/96.4\%)  & 139 & 441 & 220 & $-0.14 \pm 0.08$ & Yes \\
Sparse WR Mobile  & 71 & 456 & 273 & $-0.40 \pm 0.09$ & Yes \\
Subscale WR 1024 ($16\times$) & 113 & 558 & 129 & $-0.03 \pm 0.05$ & No\\ 
\bottomrule
\end{tabular}
\end{sc}
\end{small}
\end{center}
\vskip -0.1in
\caption{Results of A/B comparison tests between a given model and the WaveRNN-896. Each test includes 800 human ratings with grades between -3 (Much Worse Than) and 3 (Much Better Than). We collapse the counts for the different positive and negative categories.}
\label{ab}
\vskip -0.1in
\end{table*}

We propose a generation process based on \emph{subscaling}. A tensor of scale $L$ is folded into $B$ sub-tensors of scale $L/B$. The $B$ sub-tensors are generated in order, each conditioned on the previous sub-tensors. Subscaling lets us generate multiple samples at once in a batch. Since the conditioning of the generation of each sub-tensor on previous sub-tensors requires in practice only a relatively small future horizon, the generation of the next sub-tensor may start soon after the start of the generation for the previous sub-tensor. It is possible in principle -- although not necessary in practice -- to recover distant future and past dependencies beyond the horizon; the precise cost of batched sampling is then just the $B$ \emph{distant} dependencies between the samples in the current batch.
The Subscale WaveRNN is able to produce $B=16$ samples per step without loss in audio fidelity as evidenced by A/B comparison tests (Table~\ref{ab}). Batched sampling for the Subscale WaveRNN opens many orthogonal ways of increasing sampling efficiency. Even our regular Tensorflow implementation of the model achieves real-time  sampling speed on a Nvidia V100 GPU. A Fused variant of Subscale WaveRNN also gives a sampling speed of $10\times$ real time on a Nvidia P100 GPU using a slight modification of the GPU kernel for WaveRNN-896.

\section{Wave Recurrent Neural Networks}

Convolutional sequence models \cite{bytenet} achieve excellent performance in speech synthesis~\cite{tacotron}, yet their architecture tends to be deep and narrow requiring a long chain of layers to be executed for each sample. We seek an architecture that provides an equally expressive and non-linear transformation of the context, but requires a small number of operations at each step. By having a hidden state that maintains an already compressed representation of the context, an RNN is especially suitable for this purpose as it is able to combine the context with the input within a single transformation.
The overall computation in the WaveRNN is as follows (we omit biases for brevity):
\begin{align}
\label{wavegrueq}
\nonumber&\vec{x}_t = [\vec{c}_{t-1},\vec{f}_{t-1}, \vec{c}_t] \\
\nonumber&\vec{u}_t = \sigma(\vec{R}_u \vec{h}_{t-1} + \vec{I}^\star_u \vec{x}_t) \\
\nonumber&\vec{r}_t = \sigma(\vec{R}_r \vec{h}_{t-1} + \vec{I}^\star_r \vec{x}_t) \\
\nonumber&\vec{e}_t = {\tau}(\vec{r}_t \circ (\vec{R}_e \vec{h}_{t-1}) + \vec{I}^\star_e \vec{x}_t) \\
&\vec{h}_t = \vec{u}_t \circ \vec{h}_{t-1} + (1 - \vec{u}_t) \circ \vec{e}_t \\
\nonumber&\vec{y}_c, \vec{y}_f = \mbox{split}(\vec{h}_t) \\
\nonumber&P(\vec{c}_t) = \mbox{softmax}(\vec{O}_{2}\ \mbox{relu}(\vec{O}_{1} \vec{y}_c)) \\
\nonumber&P(\vec{f}_t) = \mbox{softmax}(\vec{O}_{4}\ \mbox{relu}(\vec{O}_{3} \vec{y}_f))
\end{align}

where the $\star$ indicates a masked matrix whereby the last coarse input $\vec{c}_t$ is only connected to the fine part of the states $\vec{u}_t, \vec{r}_t, \vec{e}_t$ and $\vec{h}_t$ and thus only affects the fine output $\vec{y}_f$. The coarse and fine parts $\vec{c}_t$ and $\vec{f}_t$ are encoded as scalars in $[0, 255]$ and scaled to the interval $[-1,1]$. The matrix $\vec{R}$ formed from the matrices $\vec{R}_u, \vec{R}_r, \vec{R}_e$ is computed as a single matrix-vector product to produce the contributions to all three gates $\vec{u}_t, \vec{r}_t$ and $ \vec{e}_t$ (a variant of the GRU cell as in \cite{gru, baidu_diffgraph}.) $\sigma$ and $\tau$ are the standard sigmoid and tanh non-linearities. A possible architectural variant is to have $h_t$ depend only on $x_{t-1}$ and use a fully connected layer followed by summation or concatenation to condition $f_t$ on $c_t$; we found that this version required $20\%$ more parameters and also performed 1-2 centi-nats worse.

We split the state of the RNN in two parts that predict respectively the 8 coarse (or more significant) bits $\vec{c}_t$ and the 8 fine (or least significant) bits $\vec{f}_t$ of the 16-bit audio sample (Figure~\ref{wavernn}). Each part feeds into a softmax layer over the corresponding 8 bits and the prediction of the 8 fine bits is conditioned on the 8 coarse bits. The resulting \emph{Dual Softmax} layer allows for efficient prediction of 16-bit samples
using two small output spaces ($2^8$ values each) instead of a single large output space (with $2^{16}$ values). Figure~\ref{wavernn} shows this visually. We note that it is possible to train with one softmax over all $2^{16}$ values, but that in addition to requiring significantly more parameters, memory and compute, it consistently performs 1-2 centi-nats worse.

\subsection{WaveRNN Sampling on GPU}

The above architecture reduces the number of operations $N$ that are needed for each step from $N=60$ for WaveNet with the 16-bit Discretized Logistic Mixture (DLM) output \cite{salimans} to $N=5$ for the proposed WaveRNN with the dual softmax. 
Despite the reduced number of operations $N$, a regular implementation of WaveRNN sampling 
does not directly yield a real-time or faster synthesis. On a GPU the primary hindrance is not the raw FLOPs required for sampling;
rather, the difficulties are twofold: limits on the memory bandwidth and the time that it takes to launch each of the $N$ operations. Regarding the former, a WaveRNN with a state of 896 units (WaveRNN-896) has about 3M parameters. A regular implementation of sampling that calls each \mbox{WaveRNN} operation separately in sequence for each of the 24,000 samples loads all of the WaveRNN parameters from memory into the GPU registers during each step, totalling about $3e6 \times 24e3 \times 4 = 288$ GBytes of required memory bandwidth. This is already more than a third of the memory bandwidth available in an Nvidia P100 GPU, giving by itself an upper bound of $3\times$ real time for a regular implementation of sampling.

\begin{table}[t]
\vskip 0.15in
\begin{center}
\begin{small}
\begin{sc}
\begin{tabular}{lccrr}
\toprule
Size & Sparsity \% & Type & Platform & Samples/sec \\
\midrule
512 & 95\% & 4$\times4$ & SD 835 & 29,100  \\
512 & 95\% & 4$\times4$ & SD 808 & 19,800  \\
512 & 95\% & 16$\times$1 & SD 835 & 31,400 \\
512 & 95\% & 16$\times$1 & SD 808 & 21,600 \\
\bottomrule
\end{tabular}
\end{sc}
\end{small}
\end{center}
\caption{Benchmarks for Sparse WaveRNN Mobile sampling performance executed on the widely available Snapdragon 808 and 835 mobile CPUs. The model has 1024 hidden units, $95\%$ sparsity and $4\times4$ structure sparsity. The benchmarks are based on running an equivalent computation on the mobile CPU (Section~\ref{sparse_samp}).}
\label{swrnnspeed}
\end{table}

The overhead of launching each operation separately on the GPU is even larger. 
While launching an operation on the GPU has a constant overhead of 5 microseconds, each step requires  $N=5$  such operations, which means the launch overhead alone induces an upper bound of 40,000 samples per second. For the WaveNet architecture, which requires (at least) $N=60$ operations per sample, the launch overhead induces an upper bound of 3,300 samples per second. This is without considering the time spent on the actual computation of the operations. In practice a regular implementation of sampling in e.g. Tensorflow yields, respectively, about 1600 and 170 samples per second for WaveRNN-896 and for WaveNet.

\begin{figure}
    \centering
    \includegraphics[scale=0.6]{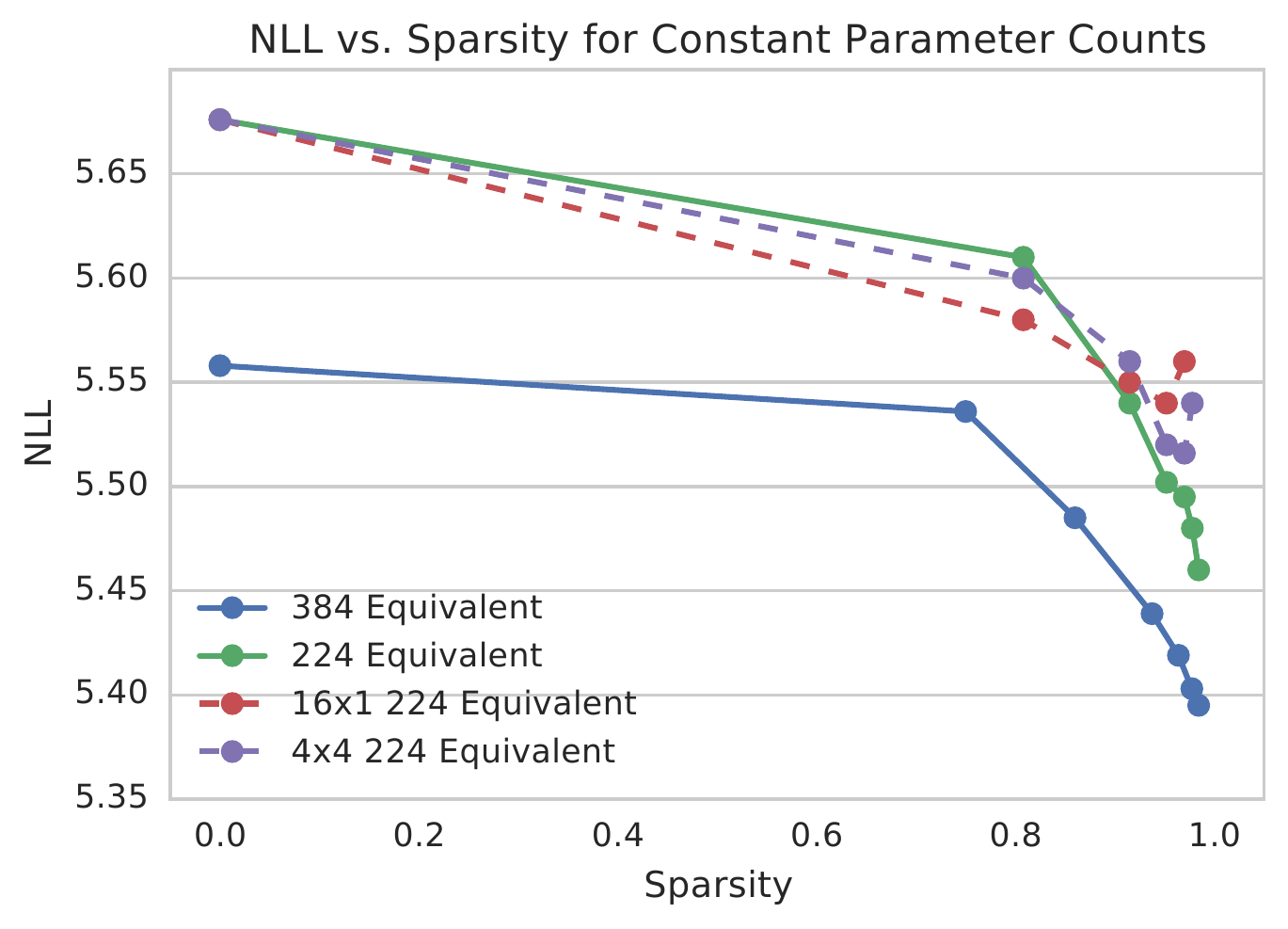}
    \vskip -0.1in
    \caption{The Sparse WaveRNNs on each curve have the same number of parameters. The Sparse WaveRNNs with structured sparsity $16\times1$ and $4\times4$ hit a point of maximum performance at a high degree of sparsity. The points of maximum performance for the unstructured Sparse WaveRNNs fall beyond the tested range.}
    \label{fig:sparsity}
\end{figure}

We reduce both of these factors by implementing the sampling procedure directly as a single persistent GPU operation. The memory bandwidth bottleneck is avoided since the parameters are loaded only once into the GPU registers at the start of sampling and persist in the registers throughout the process. This is possible because the P100 GPU has 3.67M full-precision registers that suffice to store more than 7 million half-precision parameters, i.e. more than twice as many as needed in the WaveRNN-896. The operation launch bottleneck is also avoided, since the entire sampling process for an utterance is executed as a single GPU operation.

A state size of 896 is chosen specifically to fit the P100 GPU which has 56 multi-processors.  The minimum numbers of warps that must be assigned to each multi-processor to access the full register file of the GPU is 8. If we assign each warp to a state calculation, then the state size must be a multiple of $56*8=448$ and the largest multiple that fits in the available register space is 896.

The resulting GPU kernel for WaveRNN sampling is two orders of magnitude more efficient than the regular sampling implementation, reaching 96,000 samples/second for the WaveRNN-896. The corresponding operation for WaveNet reaches 8,000 samples/second. The new overhead $d(op)$ is now given by the synchronization of the thousands of cores in the GPU~\cite{gpubarriers}, which takes just 500 nanoseconds per synchronization, instead of the 5 microseconds needed for each operation launch.

\begin{figure*}
\centering
\includegraphics[width=\textwidth]{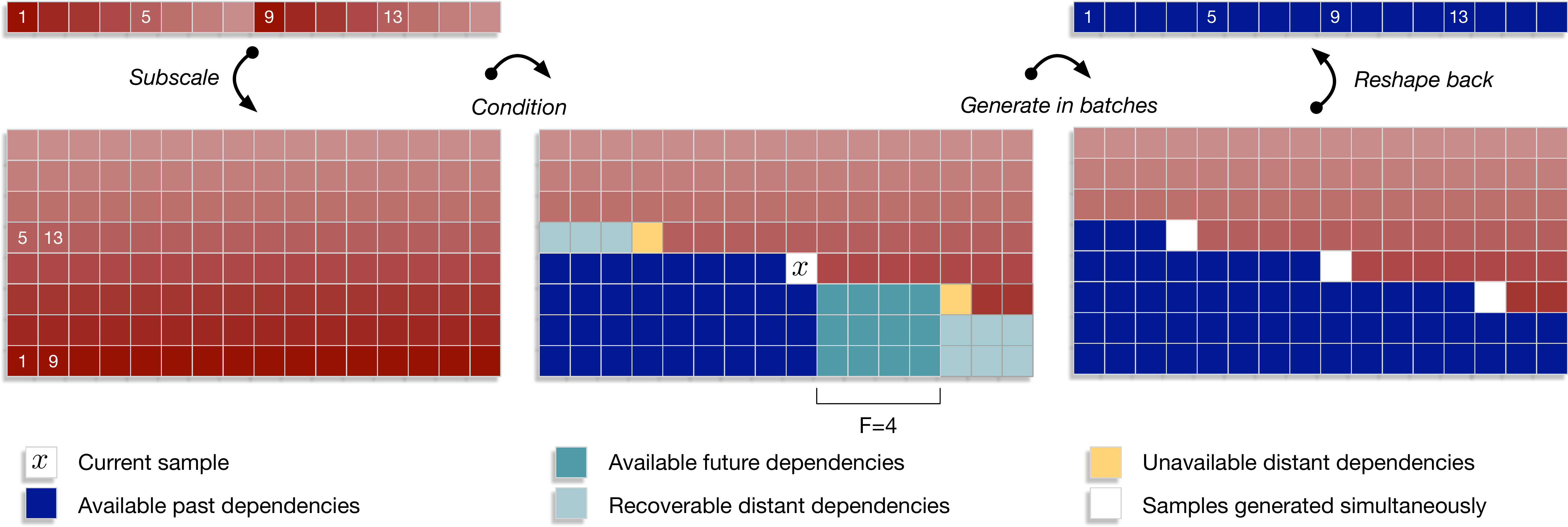}
\caption{
The dependency scheme of Subscale WaveRNN. Each box corresponds to one 16-bit sample. Subscaling first reshapes the tensor into $B$ sub-tensors of interleaving samples. Then each sub-tensor is generated conditioned on past and future samples of previously generated sub-tensors; the past horizon is unbounded, whereas the future horizon of size $F$ is tied to the receptive field of the conditioning network. Batched sampling can then be applied. The final tensor in the original scale is reconstituted from the generated sub-tensors.  }
\label{wavernn-dep}
\vskip -0.1in
\end{figure*}

\section{Sparse WaveRNN}

The WaveRNN architecture dramatically reduces the number of required operations $N$ and implementing sampling as a single GPU operation eliminates much of the original computation $c(op_i)$ and overhead $d(op_i)$ bottlenecks. 
We next present a technique for reducing directly the amount of computation $c(op_i)$ required by each operation. Decreasing the number of hidden units will reduce the amount of computation, but this comes with a significant loss in quality (Table~\ref{wavernn-results}). Instead, we reduce the number of non-zero weights in the network by sparsifying the weight matrices while retaining a large state size and respective representation capacity. This reduces $c(op_i)$ since the number of non-zero weights is directly proportional to $c(op_i)$ (Table~\ref{armperf}).

\subsection{Weight Sparsification Method}

We use a pruning scheme based on the weight magnitude that increases sparsity as training proceeds \cite{exploringsparsity,topruneornot}. We maintain a binary mask specifying the sparsity pattern of weight matrices. At the beginning of training, the weight matrices are dense.
Every 500 steps, the weights within each sparsified layer are sorted by their magnitude and the mask is updated by zeroing out $k$ weights with the smallest magnitude. The number $k$ is computed as a fraction $z$ of the total number of weights, which is gradually increased from $0$ to the target sparsity $Z$ as a function of the training step $t$: $$z = Z \left(1 - \left(1 - \frac{t - t_0}{S}\right)^3\right)$$
where $t_0$ is the step at which weight pruning begins and $S$ is the total number of pruning steps. We use $t_0=1000$, $S=200k$ and train for a total of $500k$ steps for all models. Such a scheme is practical, easy to integrate into existing models, and does not increase the training time. We sparsify the three gate matrices within the GRU cell separately.

\subsection{Structured Sparsity}

We need to encode the sparsity mask in a manner that allows for efficient computation.  The standard Compressed Sparse Row format 
uses about the same amount of storage for encoding the sparsity mask as it does for storing the parameters.  Unlike hardware-oriented approaches such as Viterbi pruning \cite{viterbi-based}, we explore structured sparsity as a means for reducing memory overhead.
The structure in the sparsity mask that we consider is in the form of non-overlapping blocks of weights which are pruned or retained together based on the average magnitude of the weights within the block.
We find that blocks of $m=16$ weights lose little performance over unstructured sparsity while reducing the amount of memory  needed for storing the sparsity pattern to $\frac{1}{m}$ of that required by an unstructured mask.
Besides rectangular $4 \times 4$ blocks that we found to work well \cite{openai_sparse,blocksparse}, we also adopt  blocks of shape $m \times 1$ that induce an even lower memory bandwidth overhead. In the case of $m\times1$ blocks one only needs to retrieve a single activation value from the hidden state to perform the dot product. This is in contrast with the square blocks where for each block one needs to retrieve $4 $ activation values from the hidden state. We report results for both $16 \times 1$ and $4 \times 4$ blocks. The benchmarks confirm the greater speed of the $16 \times 1$ blocks (Table~\ref{armperf}).

\subsection{Sparse WaveRNN Sampling on Mobile CPU}

We take advantage of the low computation and memory bandwidth required by Sparse WaveRNN to implement
matrix-vector operations necessary for
sampling on mobile CPU.
To maximize memory utilization, weights are stored in 16-bit floating point and converted to 32-bit floating point before being used in the computation. The activations and the calculations are kept in 32-bit floating point.
The low memory overhead afforded by small blocks allows the sparse matrix-vector products to match the performance of dense matrix-vector products with the same parameter count. The number of sequential matrix-vector products per second 
is thus determined almost entirely by the number of parameters in the network.



\section{Subscale WaveRNN}

We have described two ways of reducing sampling time in high-fidelity audio generation: the WaveRNN that reduces $N$ and $d(op)$ and the Sparse WaveRNN that reduces $N$ and $c(op)$. Lastly we reduce the contribution from the factor $|\vec{u}|$ in Equation \ref{latency}. This factor depends on the size of the utterance $\vec{u}$ and a direct reduction of the size of $\vec{u}$ itself (such as going from 16 to 8 bits per sample) would negatively affect audio quality. Instead, we propose a method for generating a batch of $B$ samples per step, instead of just one:
\begin{equation}
\label{latency_batch}
T(\vec{u}) = \frac{|\vec{u}|}{B} \sum_{i=1}^{N} (c(op_i^B) + d(op_i^B))
\end{equation}
In many cases, the computation time for a batch of $B$ examples, $c(op_i^B)$, grows sublinearly in the computation time of a single example $c(op_i)$ because weights are reused and spare computational capacity is available. The ability to batch samples also makes it possible to generate across multiple processors and have a reduction in total sampling time that is linear in the number of processors.

Previous work on producing more than one sample per step in sequential models has required breaking \emph{local} dependencies \cite{DBLP:conf/icml/ReedOKCWCBF17}: two nearby samples that strongly depend on each other are produced independently, possibly conditioned on other samples. We introduce a general method that allows us to trade a small  constant number of \emph{distant} past and future dependencies for the ability to generate batches of $B$ samples per step.

\begin{table}[t]
\vskip 0.15in
\begin{center}
\begin{small}
\begin{sc}
\begin{tabular}{lcccr}
\toprule
Model   & NLL & MOS  \\
\midrule
WaveNet & 5.29 & 4.51 $\pm$ 0.08 \\ \midrule
WaveRNN 224 & 5.67 & 3.73 $\pm$ 0.09 \\
WaveRNN 384 & 5.56 & 4.23 $\pm$ 0.09 \\
WaveRNN 896 & 5.42 & 4.37 $\pm$ 0.07 \\
WaveRNN 2048 & 5.33 & 4.46 $\pm$ 0.07 \\ \midrule
{Sparse WR Mobile}  & 5.52 & 4.33 $\pm$ 0.08 \\
Sparse WR 224 / 1536@97.8\% & 5.48 & 4.39 $\pm$ 0.07 \\
Sparse WR 384 / 2048@96.4\% & 5.42 & 4.48 $\pm$ 0.07 \\ \midrule
Subscale WR 1024 ($16\times$) & 5.52 & 4.30 $\pm$ 0.08\\
Subscale WR 1024 ($8\times$) & 5.46 & 4.39 $\pm$  0.06\\ 
Fused Subscale WR 896 ($2\times$) & 5.45 & 4.31 $\pm$ 0.08 \\
\bottomrule
\end{tabular}
\end{sc}
\end{small}
\end{center}
\caption{
WaveRNN NLL and MOS results on the text-to-speech benchmark. The Sparse WaveRNN Mobile model has 1024 hidden units with a $95.2\%$ sparsity ratio and 4$\times$4 blocks.
}\label{wavernn-results}
\vskip -0.2in
\end{table}

\subsection{Subscale Dependency Scheme}

From the tensor $\vec{u}$ one first extracts a set of $B$ sub-tensors that have a frequency or scale that is $B$ times smaller. Each sub-tensor corresponds to a subscale slice of $\vec{u}$ (see Figure~\ref{wavernn-dep}). If $\vec{u}$ is a 24kHz audio utterance and $B$ is 16, then each sub-tensor  corresponds to a 24/16=1.5 kHz utterance.  This is in contrast with a multi-scale scheme where the different subtensors extracted from $\vec{u}$ have increasing scales. Subscaling induces the following  ordering on the dependencies of the variables in $\vec{u}$, which is equivalent to the standard factorization of the joint:
\begin{align}
\label{reordering}
\nonumber P(\vec{u}) = \prod_{s=0}^{B} \prod_{i=0}^{|\vec{u}|/B} P\Big(u_{Bi + s} \Big| {u_{Bj + s} \mbox{ for } j < i}, \\ u_{Bk + z} \mbox{ for }  z < s \mbox{ and } k\geq 0\Big)
\end{align}
The sample $u_{Bi + s}$ for a given $(i,s)$ depends on all samples $u_{Bk + z}$ for $z<s$ and $k\geq0$. Generation of $\vec{u}$ proceeds as follows: one first generates the first sub-tensor, then the second sub-tensor conditioned on the first one, then the third sub-tensor conditioned on the previous two, etc. The Subscale WaveRNN that generates a given sub-tensor is conditioned on the future context of previous sub-tensors using a masked dilated CNN with relus and the mask applied over past connections instead of future ones.
Like the multi-scale scheme, subscale schemes are equally applicable to multi-dimensional tensors.

\subsection{Batched Sampling}

In contrast to the multi-scale scheme, subscaling makes it possible to generate $B$ samples in a single step. 
In Equation~\ref{reordering}, for values of $k > i + F$ for some \emph{future horizon} $F$, the dependencies of $u_{Bi + s}$ on future samples $u_{Bk + z}$ with $z<s$ become overwhelmingly weak (Figure~\ref{wavernn-dep}). The conditioning network itself in the Subscale WaveRNN only sees a finite and usually small number of future samples from the previous sub-tensors. The sampling of a sub-tensor can begin immediately after the first $F$ samples of the previous sub-tensor have been generated. Because the Subscale WaveRNN is shared across all sub-tensors, it is possible to batch inputs and after $B*F$ steps the total batch of the Subscale WaveRNN is $B$.  Since the value of $F$ (usually 64 or 128) is relatively small compared to the scale and length of $\vec{u}$, even for relatively large values of $B$ such as 16, the total lag of $B*F$ steps remains negligible for the total sampling latency. 
Although the conditioning network needs to be executed for each batch of samples, computing the conditioning network doesn't affect the factor $N$ of the Subscale WaveRNN because the network can be executed in parallel for a chosen number $L$ of future samples. This increases the total sampling lag by $B*L$ steps, which even for values of $L=100$ remains negligible. Due to batched sampling even our regular implementation in Tensorflow achieves just about real-time speed (24,000 samples/second) for a Subscale WaveRNN $16\times$ with 1024 hidden state units.

\subsection{Recovering Future and Past Dependencies}

Dropping distant future dependencies for $k>i+F$ allows us in principle also to recover an almost equal number of distant past dependencies. A sub-tensor $z$ that succeeds the current sub-tensor $s$ is $(z-s)(F+1)$ steps behind $s$, but leaves a trace of distant past samples. During training and sampling these distant past samples can be accessed to condition the generation of the current pass $s$. Analogously, a constant number of future distant samples beyond $i+F$ from sub-tensors previous to $s$ are also available for additional conditioning. The exact dependency scheme of using subscaling and batched sampling  includes these distant dependencies; in practice, however, choosing a larger value $F$ appears simpler than embedding the distant dependencies.

\subsection{Fused Subscale WaveRNN}

We use the scheme behind the Subscale WaveRNN to directly generate more than 16 bits per step in the WaveRNN itself. We take a Subscale WaveRNN $2\times$ model and instead of batching the 2 sub-tensors we split the hidden state of the \mbox{WaveRNN} in two parts. We then use 8 softmaxes of 4 bits each and an $F$ value of just two. The samples from the sub-tensors are given directly to the \mbox{WaveRNN} as input without using a conditioning network. The resulting Fused Subscale WaveRNN $2\times$ achieves only a small drop in the quality of the output (Table~\ref{wavernn-results}), but maps well onto the WaveRNN GPU custom operation.  
Compared to WaveRNN which runs at $4\times$ real time, this model generates 32 bits per step and requires fewer synchronizations, resulting in a sampling speed of $10\times$ real time.
We note that in contrast to the Subscale WaveRNN, because fusion requires splitting the hidden state, audio quality drops quickly for factors beyond $2\times$ in the Fused Subscale WaveRNN.

\section{Experiments}
\label{experiments}

We perform experiments on the text-to-speech synthesis task and report the quality evaluation results as well as the sampling speed of our benchmarks on the corresponding platforms.

\label{dataset}

Text-to-speech models were trained on a dataset of 44 hours of North American English speech recorded by a professional speaker \cite{parallel_wavenet}. The generation is conditioned on conventional linguistic features and predicted pitch information. All compared models synthesize raw audio at 24 kHz in 16-bit format.
The evaluation is carried out on a held-out test set where we consider three performance measures: Negative Log-Likelihood of ground-truth audio; MOS between 1 (Bad) and 5 (Excellent) of generated speech utterances according to the subjective quality evaluation by human raters; and the results of direct A/B comparison tests between pairs of models as rated subjectively by humans on a scale between -3 (Much Worse Than) and +3 (Much Better Than).

\subsection{WaveRNN Quality Evaluation \& Speed}
\label{wavernn_eval}
\label{wavernn_speed}
The WaveRNN models are trained on sequences of 960 audio samples of 16-bit each and full back-propagation-through-time is applied to the models.
Table~\ref{wavernn-results}  reports the results for various sizes of WaveRNN. The larger WaveRNNs approach the NLL performance of the 60-layer WaveNet model. A human rated A/B comparison test between WaveRNN-896 and WaveNet indicates no significant difference in the quality of the speech produced (Table~\ref{ab}). An additional A/B comparison test between WaveNet and WaveRNN-2048 also shows no significant differences. 


The persistent GPU operations that we implement are most efficient for the WaveRNN-896 model, which achieves a NLL of 5.42 and a MOS value of $4.37 \pm 0.073$. Samples are generated at 96,000 samples per second for a batch size of 1 and 39,000 samples per second for a batch size of 4.

\subsection{Sparse WaveRNN Quality Evaluation \& Speed}
\label{sparseresults}

Figure~\ref{fig:sparsity} illustrates a core point about our investigation into sparse models. We use a dense WaveRNN model with a state size of 224 as a starting point because it is the largest that could be run on many current off the shelf mobile processors.  As a second baseline we use a model with a state size of 384 that we estimate to still be out of reach for even the fastest mobile platforms, as the model would require 30 GB/sec of memory bandwidth and no current mobile platform can provide this amount. 

Figure~\ref{fig:sparsity} shows that if we fix the total parameter count -- and keep the corresponding sampling time also the same -- then as we increase the degree of sparsity and the resulting size of layers, the fidelity of the models improves. This holds up to high degrees of sparsity $>98\%$, where the state size $h$ of the models reaches 2048 hidden units. Higher sparsity monotonically implies lower NLL and in fact  higher sparsity levels have larger slopes. 
This suggests that for a given computational budget at inference time, it is much more efficient to use those parameters to sparsely connect a larger number of neurons in each layer.


\begin{table}[]
\vskip 0.15in
\begin{center}
\begin{small}
\begin{sc}
\begin{tabular}{lccccccc}
\toprule
Size & Sparse  & Type & \multicolumn{2}{c}{SD 808}  & \multicolumn{2}{c}{SD 835} \\
& \% & &  GF & MVM & GF & MVM  \\
& & &   & $\times10^3$ &  & $\times10^3$  \\
\midrule
224  & 0    & -    & 9.6 & 95.4 & 11.0 & 95.4 \\
384  & 0    & -    & 9.6 & 32.6 & 11.0 & 32.6 \\
1024 & 0    & -    & 3.8 & 1.8  & 8.0 & 3.8 \\ \midrule
512  & 80.0 & 1x1  & 2.1 & 20.1 & 3.8 & 36.5\\
1024 & 95.0 & 1x1  & 1.8 & 17.2 & 3.4 & 32.2 \\
2048 & 96.4 & 1x1  & 2.0 & 6.5  & 3.7 & 12.1 \\ \midrule
512  & 80.0 & 4x4  & 8.9 & 85.2 & 14.3 & 136.6 \\
1024 & 95.0 & 4x4  & 8.0 & 75.6 & 12.4 & 118.2 \\
2048 & 96.4 & 4x4  & 8.5 & 28.1 & 12.8 & 42.2 \\ \midrule
512  & 80.0 & 16x1 & 9.8 & 94.0 & 14.5 & 138.1\\
1024 & 95.0 & 16x1 & 9.0 & 85.5 & 13.4 & 127.4 \\
2048 & 96.4 & 16x1 & 9.0 & 30.0 & 12.6 & 41.8 \\
\bottomrule
\end{tabular}
\end{sc}
\end{small}
\end{center}
\vskip -0.1in
\caption{Performance of ARM matrix-vector multiplies (MVM) and respectively Gflops (GF) \emph{per second}, using two big cores of each of the processors Snapdragon 808 and 835. For the dense 224 and 384 kernels, higher performance is possible (11.7 Gflops/sec and 16.3 Gflops/sec respectively) with custom layouts of the dense matrix, but this is best performance we could achieve with the standard row major layout.}
\label{armperf}
\vskip -0.2in
\end{table}

\label{sparse_samp}

In Table~\ref{armperf} and Figure~\ref{fig:sparsity} we examine the impact of using block sparsity on NLL and speed, and find that $4\times4$ blocks generally yield the best NLL, but $16\times1$ blocks have a speed advantage. Surprisingly, both have better NLL than unstructured sparsity at low sparsity levels, but improve more slowly and eventually hit a minimum around 95\% while unstructured sparsity continues to improve. We did not explore even higher levels of unstructured sparsity only because investigating extremely high levels of sparsity requires starting with extremely large dense layers making training computationally intensive. Unstructured sparsity is unsurprisingly slower during inference, but depending on the quality trade offs involved in using blocks (which will likely vary from model to model), it might still be preferred.

To obtain an estimate of Sparse WaveRNN sampling speed, we benchmarked all computationally heavy operations (sparse matrix-vector multiplication and non-linearity evaluations) required for producing each audio sample, and used these measurements to derive an estimate of the sampling speed. For example, a sample from a 1024 model requires 3 multiplications of 1024$\times$1024 for the GRU gates, two multiplications of 512$\times$512 for the projection, two multiplications of 512$\times$256 for the logits and evaluating 3072 non-linearities.  We add up the time for all of these operations to estimate the upper bound on sampling performance.

We perform our benchmarks on the Snapdragon 808 (SD 808) and Snapdragon 835 (SD 835) mobile CPUs, which are widely available in mobile phones. The two big cores of the SD 808 at 1.8GHz can do 28.8 Gflops/sec and the bandwidth out of the shared L2 cache is 14.4 GB/sec.  The SD 835 is faster at 2.35GHz, with 2 cores able to do 37.6 Gflops/sec and pull 18.8 GB/sec from the cache. In practice, the achievable flops are often much lower~\cite{nexus5xgeekbench, pixel2geekbench} and around 14.4 Gflops/sec and 28.2 Gflops/sec for SD 808 and 835 respectively. These numbers suggest that both our dense and sparse implementations are close to the maximum possible performance of the processor (the limiting factor for all kernels is bandwidth and not flops).  For comparison, a modern Intel desktop CPU with AVX2 can do over 200 Gflops/sec and get over 200 GB/sec of bandwidth out of the L2 cache with only two cores.

\subsection{Subscale WaveRNN Quality Evaluation}

The conditioning network of the Subscale WaveRNN is a masked dilated 1D CNN and has ten layers, convolutional kernels of size 3, 384 convolutional channels, and 768 residual channels. The conditioning CNN has 5 stages of increasing dilation, for a total future horizon of $F=128$ blocks of 8 or 16 samples each. The Subscale WaveRNNs that we evaluate have 1024 units in their hidden state. We do not use recoverable distant dependencies.

We evaluate the model for two values of $B$, 8 and 16. The Subscale WaveRNN with $B=8$ generates 8 16-bit samples at once at each step, which corresponds to a 3 kHz signal. As shown in Table~\ref{wavernn-results}, the Subscale WaveRNN $8\times$ achieves a MOS of 4.39. This is equivalent to the MOS of the baseline WaveRNN-896 and it shows the ability of the Subscale WaveRNN $8\times$ to accurately learn the distribution under  the modified dependency scheme. We also evaluate a Subscale WaveRNN with $B=16$, which generates an interleaving signal at 1.5 kHz. As shown in Table~\ref{ab}, the audio fidelity of the Subscale WaveRNN $16\times$ is not significantly different from that of the WaveRNN-896 and, by transitivity, that of Wavenet 512 (60). This is remarkable as audio generation with sequential models can be extremely sensitive to  lost dependencies, especially local ones, and this quality result demonstrates the effectiveness of the subscale dependency scheme to preserve all the local dependencies that are the key to the high performance of sequential models.

The ability to batch computation by a factor of 8 or 16 yields a large amount of flexibility. Batching can increase throughput from a single GPU device increasing the overall sampling speed. In addition, it makes it possible to generate from multiple devices at once, where the generated bits are sent one-way and online from each device to the next. Such a setup gives in principle a linear speed-up over the sampling speed of a single device. If a single pass of Subscale WaveRNN with $B=16$ runs at $4\times$ real time on a GPU, then on a connected rack of 16 GPUs the Subscale WaveRNN $16\times$ can in principle gain an equivalent linear speed-up for a total sampling speed of $4*B=64$ times real time. Subscale WaveRNN can also be combined with Sparse WaveRNN and executed on a multi-core CPU  gaining a speed-up proportional to the number of cores available.

\vspace{-0.3cm}

\section{Conclusion}

We introduced the WaveRNN, a simple and powerful recurrent network for the sequential modeling of high fidelity audio, and we have demonstrated a high performance implementation of this model on GPUs. We have shown that large sparse models have much better quality than small dense models with the same number of parameters and we have written high performance block-sparse matrix-vector product operations to demonstrate that sampling time is proportional to parameter count. We then showed that high fidelity audio generation is now achievable on widely available low-power mobile CPUs. Finally, we introduced the subscale dependency scheme that lets sequential models generate many samples per step while preserving the output quality of the original model. The underlying ideas of the methods we introduce are not specific to audio, and the results of sparse models have implications for inference in all types of neural networks.

\bibliography{main}
\bibliographystyle{icml2018}

\end{document}